*Article*

# Review of Steady-State Electric Power Distribution System Datasets

Steffen Meinecke [1,*], Leon Thurner [2] and Martin Braun [1,2]

[1] Department of Energy Management and Power System Operation (e²n), University Kassel, Germany; steffen.meinecke@uni-kassel.de (S.M.); martin.braun@uni-kassel.de (M.B.)
[2] Fraunhofer Institute for Energy Economics and Energy System Technology (IEE), Kassel, Germany; leon.thurner@iee.fraunhofer.de (L.T.)
* Correspondence: steffen.meinecke@uni-kassel.de

**Abstract:** Publicly available grid datasets with electric steady-state equivalent circuit models are crucial for the development and comparison of a variety of power system simulation tools and algorithms. Such algorithms are essential to analyze and improve the integration of distributed energy resources (DERs) in electrical power systems. Increased penetration of DERs, new technologies, and changing regulatory frameworks require the continuous development of the grid infrastructure. As a result, the number and versatility of grid datasets, which are required in power system research, increases. Furthermore, the used grids are created by different methods and intentions. This paper gives orientation within these developments: First, a concise overview of well-known, publicly available grid datasets is provided. Second, background information on the compilation of the grid datasets, including different methods, intentions and data origins, is reviewed and characterized. Third, common terms to describe electric steady-state distribution grids, such as *representative grid* or *benchmark grid*, are assembled and reviewed. Recommendations for the use of these grid terms are made.

## 1. Introduction

The world wide paradigm shift from fossil fueled to sustainable and low-carbon energy systems brings profound changes to the way power systems are operated and planned. This is accompanied by numerous studies in the field of renewable energy source grid integration [1], such as studies to analyze distributed energy resources (DERs) hosting capacities [2–4], to analyze cost-efficient and secure grid planning for grids with a high share of DERs [5,6], or to simulate new solutions for smart grid operation [7]. Most of the simulations within these studies are based on steady-state power system analyses which require datasets of the respective grid models.

The fact that power system operators treat their grid data as confidential is a challenge for the scientific community, which relies fundamentally on the reproducibility of scientific studies [8]. To make power system research more available and comparable, a large body of openly available grid datasets, which can be used for research purposes, has been accumulated in the public domain. These datasets differ greatly with regards to aspects, such as the grid size, the applicability of different analyses, the intended use cases, the origin of the data, or the used generation methodologies. Therefore, it is a challenge for researchers to find and select appropriate grid datasets for their individual studies.

As stated in [9,10], grid datasets are often used beyond the originally intended use cases. Often, researchers are not aware of the scope of the provided datasets due to the lack of documentation. A concise overview of existing power system datasets and their intended use case and scope is therefore needed.

Power systems are constantly evolving and investigated use cases change. As a result, for many studies no suitable dataset might be available in the public domain. It is therefore also necessary to document methodologies and algorithms that can be used to generate new power system datasets.

*1.1. Related literature and state of the art*

A number of grid datasets are available in the Christie's Power Systems Test Case Archive [11] and on newer websites [12–19]. Open source power system analyzisation and optimization tools, such as MATPOWER [20] or pandapower [21], include grid datasets in the respective format.

In recent years, the articles [10,22,23] have compiled valuable information on grid data. However, an overview with comprehensive information about the variety of different grid datasets is difficult to find [22]. Specifically, what is still missing to support researchers in selecting appropriate grid datasets is an overview of the types of power system analyses for which the grid datasets are applicable.

In [10,22,23], information about intended use cases and also some about the geographical origin is collected. However, information on the methodologies used to compile the datasets is lacking, although it is also relevant for selecting appropriate grid datasets or for creating new datasets.

Due to the large amount of available datasets, they are often characterized or classified with terms such as *reference network*, *representative grid*, or *benchmark grid* in literature. However, as far as the authors are aware, there is no standard by a standardization committee, for instance IEC, ISO, IEEE, or CENELEC, which defines the meaning of these basic terms. Subsequently, the terms are not always used consistently in literature. This can lead to misunderstandings and incorrect expectations of grid datasets.

*1.2. Contribution of the paper and structure*

To close the identified gaps in review literature, this paper has three major objectives: First, it gives a concise overview of existing distribution grid datasets as an appropriate starting point for researchers (Section 2). Second, the paper presents background information of these grid datasets, such as intended use cases and grid compilation methodologies. This can help researchers, who need to select an existing or create a new grid, to understand the design and the purpose of existing grid dataset (Section 3). Third, the paper proposes a consistent nomenclature for common terms of distribution grids to facilitate clear communication between researchers (Section 4). Finally, a summary and conclusion is given in Section 5.

## 2. Available Grid Datasets

Using publicly available distribution grid data makes studies easily comparable to other work. When selecting an appropriate existing grid dataset, it is necessary to get an overview of the available grids and their properties. The resources available for researchers to get an overview of existing datasets [10,12,22,23] are enhanced by Table 1 to show which power system analyses are applicable to the grid datasets.

While this paper focuses primarily on distribution grids, four prominent transmission system datasets have been added to the grid selection. This gives an outlook for the expandability of the overview.

The overview provides information on the year in which the different grids were published, the voltage levels, and the number of buses. Since modeling of switches is an important factor for several analyses in distribution systems, the level of detail of switch information is also given. This is divided into three categories: no modeling (-), simple marking of the switchable lines (($\checkmark$)), and indication of the position of the switch between nodes and branch elements with optional annotation of the switch type (✓). Furthermore, the table provides specific information on which analysis types the grid data are suitable for. The analysis is assumed to be possible (✓), if the relevant input parameters are included in the grid dataset. For example, state estimation relies on measurement data and optimal power flow (OPF) analysis relies on information about generation costs and operational limits for the different electric elements. The availability of geographic coordinates (GIS) is stated because this are relevant for network expansion planning, for example. Finally, the overview includes whether time series data of loads and generators are given (✓), or at least an exemplary plot corresponding to the grid is drawn ((✓)).

**Table 1.** Overview of grid data properties and possible analyses types of publicly available, widely used distribution grids (**top**) and four exemplary transmission grids (**bottom**).

| Grid Datasets[a] | Year Published | Voltage Levels[b] | Number of Buses | Switch Models | Dynamic Models | OPF Analysis[c] | Reliability Analysis | State Estimation | Unbalanced Power Flow | Short Circuit Calculation | GIS Data | Time Series Data |
|---|---|---|---|---|---|---|---|---|---|---|---|---|
| ICPSs [24–27] | 1968, 1974, 1981, 1982 | N/A | 11, 13, 43 | - | - | - | - | - | - | - | - | - |
| IEEE Case 30 [28] | 1974 | HV | 30 | - | - | ✓ | - | - | - | - | - | - |
| Cinvalar's System [29] | 1988 | MV | 14 | (✓) | - | - | - | - | - | - | - | - |
| Baran's System [30] | 1989 | MV | 33 | (✓) | - | - | - | - | - | - | - | - |
| IEEE DTFs [23,31–33] | 1991, 2002, 2010 | MV | 4-123 | -, ✓ | - | T | - | - | ✓ | - | - | - |
| Salama's System [34] | 1993 | MV | 34 | - | - | (L) | - | - | - | - | - | - |
| Su's TDG [35] | 2005 | MV | 84 | (✓) | - | - | - | - | - | - | - | - |
| IEEE NEV [23,36,37] | 2008 | MV | 21[d] | - | - | T, L | - | - | ✓ | ✓ | - | - |
| CIGRE Systems [38] | 2009 | LV, MV, EHV | 13-44 | (✓) | ✓ | G, EN, T, (L)-L | - | - | (✓) | (✓) | - | (✓) |
| IEEE 8500 NTF [23,39] | 2010 | LV, MV | 8500[d] | - | - | T, L | - | - | ✓ | - | ✓ | - |
| Kerber Grids [40] | 2011 | LV | 10-386 | - | - | V, T, L | - | - | - | - | - | (✓) |
| UKGDSs [41,42] | 2011 | MV, HV | 52-413 | - | - | ✓ | - | - | - | - | - | - |
| ATLANTIDE [43–45] | 2012 | MV | 97-103 | (✓) | ✓ | ✓ | ✓ | - | - | - | - | - |
| Dickert's LVDNs [46] | 2013 | LV | 1-150 | - | - | L | - | - | - | - | - | - |
| IEEE LVNTS [23,47] | 2014 | LV, MV | 342[d] | - | - | T, (L) | - | - | ✓ | - | - | - |
| ELVTF [23,48] | 2015 | LV | 906 | - | - | T | - | - | ✓ | - | ✓ | ✓ |
| EREDNs [49] | 2016 | LV, MV | 13-6921 | - | - | V, T, L | - | - | - | - | -[e] | - |
| SimBench [50] | 2019 | LV, MV, HV, EHV | 15-380 | ✓ | - | V, G, T, L | - | ✓ | - | - | (✓), ✓ | ✓ |
| IEEE RTS [51] | 1979 | HV, EHV | 24 | - | - | C, G, T, L | ✓ | - | - | - | - | ✓ |
| IEEE Case 9 [52] | 1980 | EHV | 9 | - | ✓ | G | - | - | - | - | - | - |
| IEEJs [53] | 2000 | HV | 236-933 | ✓ | - | T, L | ✓ | - | - | - | - | - |
|  |  | EHV | 47-115 | - | ✓ | G, (L) | - | - | - | ✓ | - | (✓) |
| PEGASE Cases [54,55] | 2015 | HV, EHV | 89-13659 | - | - | ✓ | - | - | - | - | -[e] | - |
| RTE Cases [54] | 2016 | MV-EHV | 1888-6515 | - | - | ✓ | - | - | - | - | -[e] | - |

[a] Complete grid names: Ill-Conditioned Power Systems (ICPSs), IEEE Distribution Test Feeders (IEEE DTFs), Su's Taiwanese Distribution Grid (Su's TDG), IEEE Neutral-to-Earth Voltage Test Case (IEEE NEV), IEEE 8500-Node Test Feeder (IEEE 8500 NTF), United Kingdom Generic Distribution Systems (UKGDSs), Dickert's LV Distribution Networks (Dickert's LVDNs), IEEE 342-Node LV Networked Test System (IEEE LVNTS), European LV Test Feeder (ELVTF), European Representative Electricity Distribution Networks (EREDNs), IEEE Reliability Test System (IEEE RTS), Grids of the Institute of Electrical Engineers in Japan (IEEJs)

[b] EHV > 145 kV ≥ HV > 60 kV ≥ MV > 1 kV ≥ LV; N/A: No available information within the initial publications

[c] C: cost data, V: voltage limits, G: generator limits, EN: external net limits, T: transformer limits, L: line limits, (L): line types

[d] Nodes are counted, i.e. any single electrical point is counted (relevant definition at inconsistent phase systems)

[e] GIS data exist, but are not publicly available

Several original datasets are modified or enhanced by additional information to facilitate further analysis. For example, system dynamic data are available for the IEEE RTS [56,57] and OPF data has been provided for the IEEE Case 9 [58]. Since there are multiple and sometimes conflicting derivatives of the original datasets, the overview in Table 1 considers solely the data contained in the initial publication.

## 3. Compilation Process of Grid Datasets

Table 1 gives an overview of grid data properties and possible analyses types. To check the suitability of a dataset for a specific use case, it can, however, also be relevant to know how and with what intention the grid dataset was compiled. While grid models and their parameters can be specified clearly by mathematical formulas and numbers, this type of background information is more difficult to precisely communicate. Therefore, a concise overview of intended use cases, data origins and compilation methodologies of several grid datasets is difficult to find. This section provides such an overview.

*3.1. Intended Use Case*

Use cases are often the starting points of compiling grid data. The intention in generating grids can range from compiling a simple test grid to compiling grids that represent certain specialized applications or use cases. A frequently occurring use case is the compilation of a grid that is representative of a region or a specific kind of power system structure. The intention for that is to extrapolate findings from this grid to other, unknown grids of the same type.

Even though grid datasets are usually provided with some specifications on the intended use cases, the grids are often applied in different contexts than originally intended [9,10], for example in [59–62]. In this case, researchers need to decide, whether the application of the grid allows to draw valid conclusions or whether adjustments to the grids are required. To facilitate the selection of appropriate grid datasets, intended use cases are specified in the first column of Table 2 for all grids introduced in Table 1.

*3.2. Region*

The geographical origin of the data is a relevant information, since the power system layout with regard to frequency, voltage levels and phase symmetry can significantly differ in different regions of the world. Consequently, a power system dataset compiled with North American data might not be appropriate for use cases in Europe and vice versa. This is especially relevant for distribution system datasets, where layouts vary greatly between North America, Europe or Asia.

*3.3. Grid Compilation Methodology*

In this section, three common methodologies to compile power system datasets are presented and compared.

Table 2. Overview of intentions, generation methodologies and origins of publicly available grids.

| Grid Datasets | Intended Use Cases | Region[a] | Information on Methodology | Methodic Origin of Data |
|---|---|---|---|---|
| ICPSs [24–27] | ill-conditioned sample systems for power flow methods | N/A | N/A | synthetic[a] |
| IEEE Case 30 [28] | test case for optimal load flow with steady-state security | N/A | adaption of existing test case | N/A |
| Cinvalar's System [29] | illustrating the problem of switch positioning for minimum distribution grid losses | North America | N/A | synthetic[a] |
| Baran's System [30] | test system for loss reduction and load balancing via network reconfiguration | North America | N/A | N/A |
| IEEE DTFs [23,31–33] | testing of new power flow solution methods for unbalanced systems | North America | N/A | N/A |
| Salama's System [34] | application example for the VAr control problem | North America | N/A | N/A |
| Su's TDG [35] | example grid for network reconfiguration | Taiwan | N/A | real |
| IEEE NEV [23,36,37] | examining the voltage rise on the neutral conductor | North America | N/A | N/A |
| CIGRE Systems [38] | benchmark system for issues of grid operation, planning, power quality, protection, stability | North America & Europe | use case driven approach based on experts decisions | derived from real grids |
| IEEE 8500 NTF [23,39] | representative of full-size distribution system with suitable complexity | North America | N/A | derived from real grid |
| Kerber Grids [40] | estimation of photovoltaic hosting capacity in LV grids | Germany | predefined classification method | synthetic |
| UKGDSs [41,42] | representative distribution grids to test and evaluate new concepts | United Kingdom | N/A | N/A |
| ATLANTIDE [43–45] | representative distribution grids to develop and simulate predictive scenarios | Italy | clustering method | real |
| Dickert's LVDNs [46] | LV benchmark grids representative of German feeders | Germany | principal component analysis and clustering method | synthetic |
| IEEE LVNTS [23,47] | testing of solvers in highly meshed LV systems | North America | N/A | N/A |
| ELVTF [23,48] | typical test feeders | Europe | N/A | N/A |
| EREDNs [49] | large-scale distribution grids representative of European grids | Europe | greenfield reference network model | synthetic |
| SimBench [12,50] | benchmark dataset with multiple voltage levels and data of time series and study cases to compare innovative solutions of multiple use cases based on power flow analysis | Germany | use case driven approach deriving grids from available data with validating against real grids [63] | synthetic |
| IEEE RTS [51] | test or compare methods for reliability analysis | North America | N/A | N/A |
| IEEE Case 9 [52] | small test system for stability studies | North America | N/A | synthetic[a] |
| IEEJs [53] | testing of power supply restoration planning and reliability analysis algorithms | Japan | N/A | N/A |
| PEGASE Cases [54,55] | bulk power systems for load flow and stability studies development of new tools for control and operational planning of the pan-European transmission network | France Europe | N/A N/A | derived from real grids synthetic |
| RTE Cases [54] | validation of mathematical methods and tools | France | snapshots from SCADAs | real |

[a] Presumably, due to the simple grid structure
N/A: No available information within the initial publications

3.3.1. Introduction of Common Grid Compilation Methodologies

Figure 1 shows the relation between the distribution of available real grid data (top) and resulting, published grids (bottom) with regard to the three methodologies discussed below. For a clear illustration, the figure is two-dimensional. In practice, more than two parameters are usually used to describe and classify grids. It depends on the use case which parameters are suitable and, thus, label the axes of the figure.

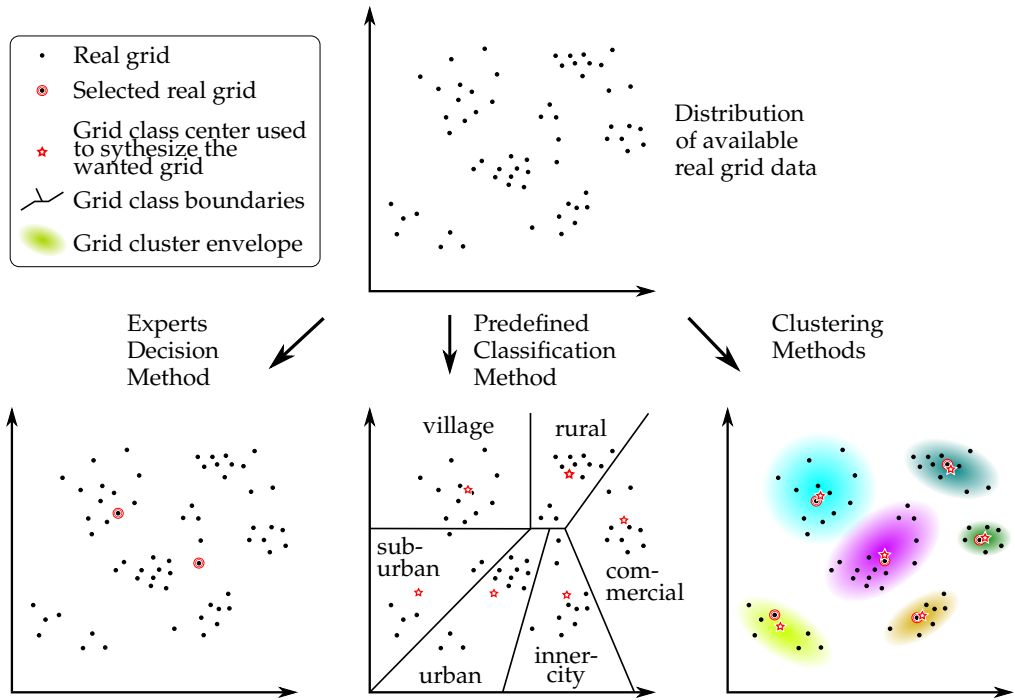

**Figure 1.** Grid selecting illustration of grid generation methods based on experts decisions (**left**), using urbanization class assumptions (**center**), without predefined grid classification (**right**).

A common method is to select grids based on expert selection decisions (see Figure 1, left). The selection is based on the data requirements derived from the intended use case. While small adjustments might be made to the grid to better fulfill the requirements, the resulting grids are of real grids origin. This method is often used for transmission systems, where the number of grids is relatively low and experts have a good overview of the characteristics of different grids. It has also been applied within the CIGRE benchmark system process [38].

A method to compile grid data based on predefined classes is shown in the center of Figure 1. In [40,64] and [65], the approach separates the grids into urbanization classes, such as *rural*, *suburban*, *urban*, or *commercial*. These classes has been defined with regard to non-electrical parameters, such as floor-space index, site occupancy index, or buildings per area. Finally, for each class, grids are synthesized using the knowledge about the parameters of the grids. Within these implementations, the approach is based on the assumption that the grids can be classified by the supply task, especially by the urbanization character.

In contrast to the before mentioned method, the classes of the method depicted on the right of Figure 1 are not defined beforehand, but compiled with mathematical clustering analyses. Multivariate, heuristic methods such as k-means or ward's method allow analyzing (dis-)similarities and appropriate groupings of the set of objects. While the resulting clusters might be interpreted as classes of grids such as *urban* or *rural*, the methodology analyzes solely the mathematical similarities. After finding a number of classes, there are two kinds of obtaining the grids, each representative of one class:

I) The best existing real grid of each class is selected, i.e. the grid with the least distance to the cluster center [66–69].

II) The parameter values of the center of each cluster are used to generate a synthetic grids with the parameters that are characteristic of the respective cluster [46,70]. As a rule, a few assumptions about the topology or parameters with missing information are required to create the grids.

3.3.2. Comparison of Methodologies

The different available methodologies each have different advantages and disadvantages, so that the appropriate method depends on the use case. Important factors in choosing a method are:

i) What is the intended use case of the dataset?
ii) Which data base can be provided for the compilation process?
iii) May data from selected grids be published or must the data be kept confidential?

The predefined classification method and the clustering methods are appropriate if the intended use case requires grids that represent a variety of real grids. If the objectives of the study already suggest certain classes of grids, the predefined classification method is suitable. For example, a study about the difference between rural and suburban power system needs to classify grids within these predefined categories. In use cases where this is not necessary, clustering methods can be used to provide the classification analyses. These are recommended over the predefined classification method, since they are based on unbiased mathematical clustering. Compared with approaches using expert knowledge, however, clustering analyses have disadvantages, such as considering causality. This can be essential depending on the intended use case. Then, either steps for selecting or adapting the grids resulting from the clustering must be added or the experts decision method needs to be applied. Notably, in [4] and [63], combinations of mathematical analyses and experts knowledge are implemented.

The requirements of the experts decision method for the data base are limited: The data considered for the decision and the data provided as resulting grids are needed. On the contrary, the predefined classification method and the clustering method require data of the investigated parameters of all investigated grids. The investigated parameters; such as rated powers of transformers, line lengths, and line types; are usually extracted from the datasets of the electric models of the real grids. Consequently, the effort for data provision and data analysis can be estimated as increasing from left to right in Figure 1. However, a selection of grids can be comprehensibly reasoned by a data analysis, whereas a selection based on the experts decision method can only be reasoned due to its applicability to the use case. This is because the data of the real grids that are not selected are usually not allowed to be disclosed.

If no data that are provided by system operators may be published, grids must be generated synthetically. Accordingly, two clustering methods has been differentiated and two different symbols has been used in Figure 1. Since a classification into these two categories is not complete, this subject is elaborated in Section 3.4.

3.4. Methodic Origin of Data

Grid datasets are classified regarding how and which data sources are used to compile the grids.
Two types of grids that data are derived from models of real grids are found: First, as with Su's TDG [35] and with the RTE cases [54], electrical parameters are, except of conversion issues, provided as they are. Second, similar to the IEEE 8500 NTF [39], the PEGASE cases [55] and the CIGRE systems [38], grid datasets are published with reference to modifications. The kind and extent of the modifications differ. As with the IEEE 8500 NTF, some elements of the grids may be changed and added, or, as with the CIGRE systems, the grids may be reduced in size.

Different types of synthetic generation of grids exist. For several studies, e.g. [40,46,70,71], grids are synthesized by filling assumed topologies with values of parameters from data analyses. In other sudies, such as [49,72–74], greenfield planning approaches are implemented. In contrast, structures like the IEEE case 9 [52] or the systems of Cinvalar [29] and Baran [30] appear simpler and could be constructed manually.

Besides grids that are compiled in a synthetic way and grids that are derived from models of real grids, different hybrid methods are possible. Differences between adapting real grids and creating synthetic grids that are gradually adapted to real networks, can be difficult to discern.

Unfortunately, several grid documentations do not specify information on the methodology and the origin of data. Especially in case of simple grids for testing, this is often skipped. Similarly, information on origins of real grid data may be omitted due to privacy concerns. As a result, some information remains unclear in Table 2 (N/A).

## 4. Terminology to Characterize Distribution Grid Datasets

As the previous section has shown, properly describing the methodology and the intended use case of grid datasets can be a complex task. Therefore, researchers often use short and succinct terms, such as *reference grid*, *synthetic grid*, or *test case*, to describe grid datasets. While this can facilitate the communication, it can also lead to misunderstandings if the terms are not clearly defined. This section reviews the use of common grid terms in the literature while considering the intrinsic meanings of the terms[1]. Moreover, a recommendation for the terminology is provided.

*4.1. Review of Grid Term Nomenclature in Literature*

4.1.1. Synthetic Grids

Grids are called *synthetic* to describe the origin of the data, e.g. in [11]. These grids are neither models of real grids nor directly derived from such. They are artificially created, for example by green field methods [74]. In [74] and [71], a number of synthetic grids are generated to achieve study results with validity. That is because simulation results can be more relevant if the algorithms run with several (types of) grids. Similarly, a large number of grids can be used to extrapolate results to real grid areas.

4.1.2. Example & Test Grids

Various research projects require grid data to exemplify or validate case studies. Well known example data are the IEEE Case 9 [52], the systems of Baran [30], Cinvalar [29], and Salama [34] as well as the ICPSs [24–27]. Real grids [75,76] and synthetically generated grids [74,77] are both named *test network*. Likewise, the number of buses varies widely depending on the use case [58]. Often, the dataset qualification for more than one use case is not considered since the test case creation have subordinate priority compared to the focus of the study.

4.1.3. Benchmark Grids

*Benchmarking* does not originate from the field of electrical power supply but from testing and comparing the performance of business processes or software tools based on trusted procedures or datasets [22]. The IEEE test feeders are called test cases or test feeders, although they are intended to serve a benchmark for different algorithms, such as unbalanced power flow [31–33][36,37], calculation of full-size distribution systems [39][48], or handling of highly meshed LV grids [47]. The CIGRE systems [38], Dickert's LVDNs [46], and the transmission grid presented in [78] themselves are named benchmark networks while having the same intention to be appropriate to be used as a dataset to benchmark algorithms and methods.

---

[1] The discourse on common grid terms is about the terms describing the grid rather than the terms *system*, *network*, *grid*, or *case*. These four terms are considered as synonyms and are applied in common usage in this paper.

Since system operators of several countries are regulated and receive incentives for efficiency, grid planning and operation management is often viewed from a financial perspective. This lead to a second understanding: A grid with which other grids are to be compared financially is named benchmark grid or, as mentioned in Section 4.1.7, reference network [79]. However, usually the process of comparing the performance of system operators, which is subject to some challenges, is called benchmarking rather than the network itself [80].

4.1.4. Representative Grids

*Representative* is used to express a relation of a grid to real grids. Comparing algorithms gets more convincing by performing the algorithms on grids with reference to reality, i.e. on representative grids [38,45]. Furthermore, representative grids are used to elaborate technical conclusions, recommendations, and estimative projections about real grids [40,71,81]. Often several representative grids are created, each representing a different class. These classes of grids could be a subset of all grids distinguished between geographic aspacts, urbanization characteristics, or electrcal parameters, e.g. *coastal grids*, *rural grids*, or *grids with long lines*. With these findings, the methods using predefined classification and clustering, depicted in Figure 1, clearly belong to representative grids.

4.1.5. Generic Grids

The intrinsic meaning of the term *generic* virtually corresponds to *general* or *universal*. Thus, generic grids should be characteristic of (a class of) grids to bring a large number of grids together.

In [82], for instance, the generic distribution grids denote several grids of different types, generated with varying parametrization. In [83], a specific system, which is intended to be particularly suitable for testing dynamic wind studies, is introduced. Here the proposed parameters of the dynamic models are open for modifications while the steady-state parameters are intended to be fixed.

There are also term usages that do not fit the intrinsic meaning of the word. For example, in [84] *generic* is referred to a grid which is derived from a real grid to allow analyzing algorithms for DER integration. The data and intended use cases of UKGDSs correspond more closely to representative grids than to the meaning of generic. This is not resolved in the referring papers [85] and [86].

4.1.6. Typical Grids

Parameters with the most frequent occurrence are described as typical. Composing these parameters, a typical grid can be formed. Since grids named typical are also described as representative [87] or generic [85], the difference seems small or unclear. In [88], the IEEE 13-Node Test Feeder [48] is also named typical. However, since this very small grid is generated to test common features of distribution analysis software and originally named as test feeder, it is more closely related to the other IEEE test feeders than to other grids named typical.

4.1.7. Reference Grids

The term *reference* grid is also used differently. To conclude these understandings, it is used as:

a) Synonym for representative grids [44,45,76,87]
b) Synthetic network, planned optimally from greenfield [49,73,79]
c) Simplified test case [70,89]
d) Best or worst case grid (to compare to); derived from representative grids by optimal choice of variable parameters [90]

*4.2. Recommended Terminology*

The previous section showed that several terms are used inconsistently throughout the literature. To eliminate ambiguities and improve the communication in scientific language, a terminology is proposed in Table 3.

Table 3. Recommended terminology of common used grid terms

| Term | Recommended Usage |
| --- | --- |
| Synthetic Grid | Grid that either do not model a real grid or that is not obtained by simplifying or modifying models of a real grid. |
| Example & Test Grid | Grid that is simply created and used for basic testing, validation, or demonstration of only one issue. Transferring quantitative conclusions from this type of grid to conditions in real grids is doubtful. |
| Benchmark Grid | Grid that is used to compare the efficiency or validity of algorithms. When using a benchmark grid, the object of investigation is the algorithm rather than the grid itself. |
| Representative Grid | Grid that is created or selected to be considered instead of a number of grids. Since one grid can hardly be representative for all grids, there are usually multiple representative grids to cover different clusters of similar grids. |
| Generic Grid | Grid with variable parameters that allow to synthesize different grids through parametrization. While representative grids use multiple grids with fixed parameters to represent different states of grids, generic grids cover multiple states through parameter variation of one grid. |
| Typical Grid | Grid with common parameters. While representative grids intend to represent a wide range of possible grids, a typical grid claim solely to cover a common or normal grid type, so that outliers and extreme cases have little or no influence on a typical grid. |
| Reference Grid | Grid that is optimal with regard to a specific criterion, such as cost-optimality. |

To exemplify the proposed terminology, in Table 4 the widespread grid datasets introduced in Table 1 are assigned to the discussed grid terms from the steady-state power flow perspective. A distinction is made between a well-suited term (✓), a partially fitting term that does not correspond to the primary focus of the original activity generating the grid dataset ((✓)), and a term that does not correspond to the recommended terminology (-). As with Table 2, information is missing to assign the term *synthetic* to every grid (N/A).

Table 4. Application of the recommended terminology to the well-known grids: well-suited terms (✓), partially fitting terms ((✓)), inappropriate terms (-), missing information (N/A).

| Grids | Synthetic | Example/Test | Benchmark | Representative | Generic | Typical | Reference |
| --- | --- | --- | --- | --- | --- | --- | --- |
| ICPSs [24–27] | ✓ [a] | ✓ | - | - | - | - | - |
| IEEE Case 30 [28] | - | - | ✓ | - | - | - | - |
| Cinvalar's System [29] | ✓ [a] | ✓ | - | - | - | - | - |
| Baran's System [30] | N/A | ✓ | - | - | - | - | - |
| IEEE DTFs [23,31–33] | N/A | (✓) | ✓ | - | - | - | - |
| Salama's System [34] | N/A | ✓ | - | - | - | - | - |
| Su's TDG [35] | - | (✓) | ✓ | - | - | - | - |
| IEEE NEV [23,36,37] | N/A | - | ✓ | - | - | - | - |
| CIGRE Systems [38] | - | - | ✓ | - | - | (✓) | - |
| IEEE 8500 NTF [23,39] | - | - | ✓ | - | - | - | - |
| Kerber Grids [40] | - | ✓ | - | ✓ | - | - | - |
| UKGDSs [41,42] | N/A | - | ✓ | (✓) | - | - | - |
| ATLANTIDE [43–45] | - | - | (✓) | ✓ | - | - | - |
| Dickert's LVDNs [46] | (✓) | - | ✓ | ✓ | (✓) | - | - |
| IEEE LVNTS [23,47] | N/A | - | ✓ | - | - | - | - |
| ELVTF [23,48] | N/A | - | ✓ | - | - | (✓) | - |
| EREDNs [49] | ✓ | - | - | (✓) | - | (✓) | ✓ |
| SimBench [50] | (✓) | - | ✓ | (✓) | - | - | - |
| IEEE RTS [51] | N/A | - | ✓ | - | - | - | - |
| IEEE Case 9 [52] | ✓ [a] | ✓ | - | - | - | - | - |
| IEEJs [53] | N/A | - | ✓ | - | - | - | - |
| PEGASE Cases [54,55] | - | - | ✓ | - | - | - | - |
| RTE Cases [54] | - | - | ✓ | - | - | - | - |

[a] Presumably, due to the simple grid structure

Baran's, Cinvalar's, and Salama's System as well as the IEEE Case 9 are classified as example/test cases in Table 4, although they are used frequently as benchmark grids nowadays. But as stated in Section 2, the intentions of the initial publications are considered here. These do not indicate that significant effort was spent in compiling the grids or that they should be appropriate for a further use case.

It should be noted that the intrinsic meanings of the grid terms and, thus, the recommendations do not address all types of information, mentioned in Section 3, at the same time. For example, *synthetic* specifies the origin of data, whereas *benchmark* expresses the intention to be used as a database to compare algorithms. As a result, terms might also be combined and grids are assigned to multiple terms in Table 4. The EREDNs, for instance, can be classified as synthetic reference grid, since they are synthetically created and optimally planned by a greenfield planning approach.

Grids can be applied in several ways. Hence, different terms may be appropriate depending on the context. For example, a grid that was intended to be a generic grid to derive scientific conclusions about grid stability, may also be used as a benchmark grid to compare the performance of two optimization algorithm.

## 5. Conclusion

Numerous distribution grid datasets are available for power system research in the public domain. This paper provides an overview of well-known and widespread publicly available grid datasets. This includes fundamental information on each dataset; descriptive information on intention, methodology, and data origin; and information on the applicability of the grid data for different power system analyses. This overview can help researchers to select appropriate grid data for their studies. As a consequence, less supplementary data and assumptions need to be added, which would take time and could inhibit transparency and comparability.

New challenges in power system operation and planning, such as the continuous increase of DER penetration in many power systems worldwide, require periodically new or modified grid datasets. Therefore, three common methodologies for creating grid datasets are discussed to facilitate studies in the generation of new grid datasets.

Short descriptive terms are common to inform about the type of grid datasets. However, these terms, such as *reference network*, *representative grid*, or *benchmark grid*, are often used inconsistently. Therefore, the usage of common grid terms in the literature is reviewed in this paper. Regarding that and the intrinsic meanings of the terms, recommendations for grid term usage are provided to improve scientific communication on steady-state electric power distribution systems. In this way, the proposed terminology can be a valuable first step for future standardization activities. The terminology is exemplified by assigning the defined terms to the reviewed grid datasets.

**Author Contributions:** conceptualization, methodology, validation, writing–original draft preparation, writing–review and editing, visualization: S.M. and L.T.; formal analysis, investigation: S.M.; supervision, project administration, funding acquisition: M.B.

**Funding:** This work was supported by the German Federal Ministry for Economic Affairs and Energy, and the Projektträger Jülich GmbH (PTJ) within the framework of the project SimBench (FKZ: 0325917A). The authors are solely responsible for the content of this publication.

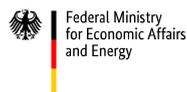

**Conflicts of Interest:** The authors declare no conflict of interest. The funders had no role in the design of the study; in the collection, analyses, or interpretation of data; in the writing of the manuscript, or in the decision to publish the results.